\documentclass{article}

\usepackage{arxiv}

\usepackage[utf8]{inputenc} 
\usepackage[T1]{fontenc}    
\usepackage{hyperref}       
\usepackage{url}            
\usepackage{booktabs}       
\usepackage{amsfonts}       
\usepackage{amsmath}
\usepackage{nicefrac}       
\usepackage{microtype}      
\usepackage{lipsum}		
\usepackage{graphicx}
\usepackage{natbib}
\usepackage{doi}
\usepackage{footmisc}
\usepackage{bbold}
\usepackage{listings}

\newcommand{\deltaB}{\boldsymbol\delta}
\newcommand{\DeltaB}{\boldsymbol\Delta}
\newcommand{\rhoB}{\boldsymbol\rho}
\newcommand{\eB}{\boldsymbol{e}}
\newcommand{\wB}{\boldsymbol{w}}
\newcommand{\NB}{\boldsymbol{N}}
\newcommand{\phiB}{\boldsymbol\phi}
\newcommand{\PhiB}{\boldsymbol\Phi}

\title{Incertus.jl – The Julia Lego Blocks for Randomized Clinical Trial Designs}

\author{ \href{https://orcid.org/0000-0003-2997-8566}{\includegraphics[scale=0.06]{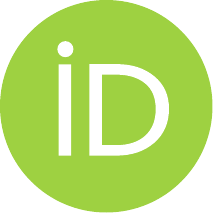}\hspace{1mm}Yevgen Ryeznik}\\
	Department of Pharmacy\\
	Uppsala University\\
	Uppsala, Sweden \\
	\texttt{yevgen.ryeznik@farmaci.uu.se} \\
	\And
	\href{https://orcid.org/0000-0002-1626-2588}{\includegraphics[scale=0.06]{orcid.pdf}\hspace{1mm}Oleksandr Sverdlov} \\
	Novartis Pharmaceuticals Corporation\\
	East Hannover, NJ, USA\\
	\texttt{alex.sverdlov@novartis.com} \\
}

\renewcommand{\shorttitle}{\textit{arXiv} Template}

\hypersetup{
pdftitle={A template for the arxiv style},
pdfsubject={q-bio.NC, q-bio.QM},
pdfauthor={David S.~Hippocampus, Elias D.~Striatum},
pdfkeywords={First keyword, Second keyword, More},
}

\begin{document}
\maketitle

\begin{abstract}
In this paper, we present Insertus.jl, the Julia package that can help the user generate a randomization sequence of a given length for a multi-arm trial with a pre-specified target allocation ratio and assess the operating characteristics of the chosen randomization method through Monte Carlo simulations. The developed package is computationally efficient, and it can be invoked in R. Furthermore, the package is open-ended -- it can flexibly accommodate new randomization procedures and evaluate their statistical properties via simulation. It may be also helpful for validating other randomization methods for which software is not readily available. In summary, Insertus.jl can be used as ``Lego Blocks'' to construct a fit-for-purpose randomization procedure for a given clinical trial design. 
\end{abstract}


\section{Introduction}
Randomization in the context of a randomized controlled trial (RCT) refers to a process of sequentially assigning eligible participants to different treatment groups through a process governed by chance \citep{Wittes2005}. Randomization helps to minimize experimental bias, ensure the comparability of treatment groups at baseline, and allow for a precise and valid treatment comparison \citep{RosenbergerLachin2015}. While randomization is an important and desirable element in studies across the entire research and development continuum, it is essential in confirmatory RCTs where the goal is to test some pre-specified clinical research hypothesis \citep{Green2002}. 

A classical RCT design is a 1:1, parallel-group, randomized, double-blind, placebo-controlled study \citep{Friedman2015}. However, modern RCTs frequently utilize multi-arm parallel group designs with equal or unequal randomization ratios. Some examples of such study designs are randomized phase 2 dose range finding studies \citep{Bretz2008}, phase 2/3 seamless designs \citep{Schmidli2006}, platform trials to evaluate multiple experimental treatments in a perpetual manner \citep{APT2019}, amongst others. The choice of a randomization method to implement the target allocation ratio for a given sample size and other trial parameters requires careful consideration. There is no shortage of randomization designs with established statistical properties \citep{RosenbergerLachin2015}; however, choosing one that is fit for purpose may be a challenging task. Some authors developed systematic approaches for the choice of an adequate randomization procedure in the context of a 1:1 RCT \citep{Hilgers2017, Berger2021, Sverdlov2024}.

The evaluation of statistical properties of randomization designs can be done both theoretically and using simulations. For 1:1 RCTs, one useful statistical software package for randomization is randomizeR \citep{Uschner2018}. For more complex settings, such as multi-arm RCTs, there is a dearth of statistical software that would enable the implementation of a chosen randomization method. Several recent references provide relevant simulation evidence on randomization in three-arm or four-arm trials \citep{RyeznikSverdlov2018, Ryeznik2018a, Ryeznik2018b}. Given the increased popularity and use of multi-arm clinical trial designs, there is a strong need for a flexible software package that enables fast and efficient generation of treatment randomization sequences for a multi-arm trial with a pre-specified target allocation ratio and enables the investigation of statistical properties of the chosen randomization method(s) through Monte Carlo simulations.

In this paper, we present Insertus.jl, the Julia package that can help clinical trial researchers to address the aforementioned goals. The highlights of Insertus.jl are:
\begin{itemize}
    \item It can be used to generate randomization sequences using a chosen method for a two-arm or a multi-arm trial with equal or unequal allocation ratios.
    \item It enables fast and efficient Monte Carlo simulation studies to compare operating characteristics of various randomization designs for the chosen trial parameters.
    \item The package is open-ended – it can flexibly integrate new procedures and evaluate their statistical properties via simulation. It can be also used as a tool to validate other randomization methods for which software is not readily available.
\end{itemize}

Section~\ref{Sec2} provides the relevant statistical background and describes the randomization methods implemented in Incertus.jl. Section 3 describes the Julia functions and presents some illustrative examples. Section 4 provides a summary and describes some potential extensions of the package.

\section{Methods}\label{Sec2}
Consider a parallel-group, randomized clinical trial to compare the effects of $K\geq 2$ treatments. Let $w_1 : w_2 : \ldots : w_K$ denote the target allocation ratio for the trial, where $w_i$'s are some positive, not necessarily equal numbers. Frequently, $w_i$'s are chosen to be integers with the common divisor of 1. For example, in a two-arm trial, $1{:}1$ allocation means that each treatment group should have an equal number of patients, whereas $2{:}1$ allocation means the first group should have twice as many patients as the second group. In some trials, an investigator may pursue a non-integer (irrational-valued) allocation ratio, e.g., the square root allocation $\sqrt{2}{:}1{:}1$ which is optimal for inference on pairwise comparisons involving treatment 1 as the reference in a three-arm trial \citep{Fleiss1986}.

Let $\rho_k=w_k/\sum_{i=1}^Kw_i$ denote the target allocation proportion for the $k$th treatment group. In our examples, $\rho_1 = \rho_2 = 1/2$ for 1:1 allocation; $\rho_1 = 2/3$, $\rho_1 = 1/3$ for 2:1 allocation; and $\rho_1=\frac{\sqrt{2}}{\sqrt{2}+2}$, $\rho_2=\rho_3=\frac{1}{\sqrt{2}+2}$ for $\sqrt{2}{:}1{:}1$ allocation. We will use the notation $\rho_1{:}\rho_2{:}\ldots{:}\rho_K$, where $\rho_i\in(0,1)$ and $\sum_{i=1}^K\rho_i=1$ to refer to any fixed, pre-specified target allocation proportions for a $K \ge 2$-arm trial.

In a clinical trial, $n$ eligible patients are enrolled sequentially and must be randomly assigned to $K$ treatments in the desired allocation ratio to achieve the sample sizes per group as $n_k \approx n\rho_k$
with $\sum_{k=1}^Kn_k=n$. For the $j$th patient in the sequence ($j=1,\ldots,n$), let $\deltaB_j=(\delta_{j1},\ldots,\delta_{jK})^T$ denote 
a $K\times 1$ random vector of treatment assignment such that only one entry of $\deltaB_j$ is equal to 1, and the rest are zeros. For example, $\delta_{jk} = 1$ and $\delta_{jk'} = 0$ for $k'\ne k$ indicates that the $j$th patient is assigned to the $k$th treatment group. This can be also written as $\deltaB_j=\eB_k$, where $\eB_k$ ($k=1,\ldots K$) is a unit vector of zeros with a 1 in the $k$th position. Let $\DeltaB_n=(\deltaB_1,\deltaB_2,\ldots,\deltaB_n)^T$ denote the $n\times K$ matrix of treatment assignments. Note that for any $j\ge 1$, $N_k(j)=\sum_{i=1}^j\delta_{ik}$ is the sample size for the $k$th treatment group after $j$ allocation steps. In particular, $N_k(n)$ is the final sample size for the $k$th group, with $\sum_{k=1}^KN_k(n)=n$.

A \emph{restricted} randomization procedure for a $K$-arm trial with the target allocation ratio $\rho_1{:}\rho_2{:}\ldots{:}\rho_K$ is defined by specifying the $j$th patient’s ($j = 1, \ldots, n$) conditional randomization probabilities, $\phi_{jk} = \Pr(\textrm{Patient}\;j\;\textrm{is assigned to treatment}\;k|\deltaB_1,\ldots,\deltaB_{j-1})$, as follows:
\begin{equation}\label{Eq1}
\phi_{1k}=\rho_k\quad\textrm{and}\quad\phi_{jk}=\Pr(\deltaB_j=\eB_k|\deltaB_1,\ldots,\deltaB_{j-1}),\;j\ge 2\;\textrm{and}\;k=1,\ldots,K.
\end{equation}

Equation (\ref{Eq1}) implies that the first patient in the sequence is randomized to treatments with probabilities $(\rho_1,\ldots,\rho_K)$, and thereafter the $j$th patient ($j\ge 2$) is randomized to treatments with 
probabilities ($\phi_{j1},\ldots,\phi_{jK}$) where $0\le \phi_{jk}\le 1$, $\sum_{k=1}^K\phi_{jk}=1$ and $\phi_{jk}$ is the conditional probability of assigning the $j$th patient to treatment $k$ given the past assignments $\deltaB_1,\ldots,\deltaB_{j-1}$. Denote $\phiB_j=(\phi_{j1},\ldots,\phi_{jK})^T$ the $K\times 1$ vector and $\PhiB_n=(\phiB_1,\phiB_2,\ldots,\phiB_K)^T$  the $n\times K$ matrix of conditional randomization probabilities for $n$ patients in the trial.

The choice of $\phiB_j$’s determines the probabilistic structure of a randomization procedure and its statistical properties. Some important and desirable features of a randomization procedure are:
\begin{itemize}
    \item[(a)] It should have a high degree of randomness, with very few (if any) deterministic assignments, to minimize the predictability and the potential for selection bias.
    \item[(b)] It should have a high degree of balance, i.e., at any allocation step, the sample treatment numbers should be close to their targeted values: $N_k(j)\approx j\rho_k$ for $j=1,\ldots,n$ and $k=1,\ldots,K$.
    \item[(c)] The unconditional probability of assigning treatment $k$ and the $j$th allocation step should be equal to the target allocation proportion for this treatment (allocation ratio preserving (ARP) property)\citep{KuznetsovaTymofyeyev2012}:
    \begin{equation}\label{Eq2}
        E(\phi_{jk}) = \rho_k\;\;\textrm{for}\;j=1,\ldots,n\;\textrm{and}\;k=1,\ldots,K.
    \end{equation}
\end{itemize}

\subsection{Two-arm equal randomization}\label{Sec2.1}
Consider a $1{:}1$ RCT to compare the effects of an experimental treatment (E) vs. control (C). In this case, $\deltaB_j=(1,0)^T$ if the $j^\text{th}$ patient is assigned to E, $\deltaB_j=(0,1)^T$ if the $j^\text{th}$ patient is assigned to C, and it is sufficient to consider a single indicator variable: $\delta_j=1$ (or $0$), if the $j^\text{th}$ patient's assigned treatment is E (or C). Let $N_E(j)=\sum_{i=1}^j\delta_i$ and $N_C(j)=j-N_E(j)$ denote the sample sizes for treatments E and C, respectively, after $j$ allocation steps, and let $D(j)=N_E (j)-N_C(j)$ denote the treatment imbalance after $j$ steps. Note that for every $j=1,\ldots, n$, $D(j)$ is a random variable whose probability distribution is determined by the implemented randomization procedure.

\subsubsection{Randomization procedures for 1:1 RCTs}\label{Sec2.1.1}
The simplest randomization procedure for a $1{:}1$ RCT is \emph{complete randomization}: each subject's treatment assignment is made independently of other assignments by a flip of a fair coin, i.e., $\Pr(\delta_j = 1) = 0.5$ for $j=1,\ldots,n$ \citep{Lachin1988}. While such an approach may be reasonable for large sample sizes, it has a non-negligible probability of treatment imbalance when the sample size is small. 

In practice, most of $1{:}1$ RCTs utilize \emph{restricted} randomization \citep{Pocock1979}. For the $j$th patient, let $\phiB_j=(\phi_j,1-\phi_j)^T$
be a vector of conditional randomization probabilities for treatments E and C. It is sufficient to consider only the first component, $\phi_j \in [0,1]$, and define a restricted randomization procedure as \citep{RosenbergerLachin2015}):
\begin{equation}
\phi_1=0.5\;\;\textrm{and}\;\;\phi_j=\Pr(\delta_j=1|\delta_1,\ldots,\delta_{j-1}),\;j\ge 2.
\end{equation}

Below we list the randomization procedures for a 1:1 RCT implemented in the current version of Incertus.jl. These procedures can be grouped into several classes depending on their probabilistic structure.\footnote{For a given procedure, there may be parameter(s) that determine a tradeoff between treatment balance and allocation randomness.} 
\begin{itemize}
    \item Complete Randomization Design (CRD): every treatment assignment is made with probability 0.5, irrespective of the previous assignments \citep{Lachin1988}.
    \item Procedures that force strict final balance ($n/2$ E’s and $n/2$ C's): Random Allocation Rule (Rand) and Truncated Binomial Design (TBD) \citep{BlackwellHodges1957}.
    \item A procedure that forces strict balance after $2b$ allocation steps, where $b$ is some small positive integer: Permuted Block Design (PBD) \citep{MattsLachin1988}.
    \item Maximum Tolerated Imbalance (MTI) procedures that control imbalance within pre-specified limits ($\pm b$) with fewer restrictions on randomization compared to PBD: Big Stick Design (BSD) \citep{SoaresWu1983}; Biased Coin Design With Imbalance Tolerance (BCDWIT) \cite{Chen1999}; Ehrenfest Urn Design (EUD) \citep{Chen2000}; Block Urn Design (BUD) \citep{ZhaoWeng2011}.
    \item Biased Coin Designs (BCDs) with the allocation function of the form $\phi_{j+1}=F(D(j))$, where $F:\mathbb{Z}\rightarrow[0,1]$ is a non-increasing, symmetric around zero function defined on the set of integers such that $F(D(j))+F(-D(j))=1$.  One notable procedure of this class is Efron’s BCD with $F_p(x)=0.5+sign(x)(0.5-p)$, where $sign(x)$ returns the value $-1$, 0, 1 if $x$ is negative, zero, or positive, respectively, and $p\in (0.5,1]$ is the fixed coin bias probability \citep{Efron1971}. Another procedure of this class is the Adjustable Biased Coin Design (ABCD) \citep{BAG2004} with the allocation function $F_a(x)=\{1+x^a\}^{-1}$, where $a\ge 0$ is a user-specified constant ($a=0$ corresponds to CRD and larger values of $a$ correspond to more balanced and less random procedures).
    \item Generalized Biased Coin Designs (GBCDs) \citep{Wei1978, Smith1984} with the allocation function of the form 
    $\phi_{j+1} = f\left(\frac{D(j)}{j}\right)$, 
    where $f:[-1,1]\rightarrow[0,1]$
    is a continuous, differentiable at 0, non-increasing function such that $f(x)+f(-x)=1$. 
    We consider a family of functions proposed by \cite{Smith1984}: $f_\rho(x)=\frac{(1-x)^\rho}{(1-x)^\rho+(1+x)^\rho}$, where $\rho\ge 0$ is a user-specified constant ($\rho=0$ corresponds to CRD and larger values of $a$ correspond to more balanced and less random procedures).
    \item Bayesian Biased Coin Designs (BBCDs) \citep{Ball1993} with the allocation rule
    \begin{equation}\label{Eq4}
        \phi_1=0.5;\;\;\phi_2=\mathbb{1}\left\{\delta_1=0\right\};\;\; \phi_{j+1}=\frac{\left\{1+\frac{N_C(j)}{jN_E(j)}\right\}^{1/\gamma}}{\left\{1+\frac{N_C(j)}{jN_E(j)}\right\}^{1/\gamma} + \left\{1+\frac{N_E(j)}{jN_C(j)}\right\}^{1/\gamma}},\;j\ge 2.
    \end{equation}
    In (\ref{Eq4}), $\mathbb{1}\left\{\delta_1=0\right\}=1$ (or 0), if the first treatment assignment was C (or E), and $\gamma > 0$ is a user-specified constant (smaller values of $\gamma$ correspond to more balanced and less random procedures).
\end{itemize}

Note that for all aforementioned 1:1 randomization procedures, the conditional randomization probability for an incoming patient can be analytically expressed as a function of the current
treatment sample sizes, i.e., we have $\phi_{j+1} = \Pr(\delta_{j+1} = 1|N_E(j), N_C(j))$. This is referred to as \emph{Markov property} \citep{Zhao2018}. However, there are 1:1 randomization procedures that do not have Markov property; e.g., the maximal procedure \citep{Berger2003}. Such procedures require the full knowledge of the completed allocation path (not just the numbers $N_E(j)$ and $N_C(j)$) to determine the treatment assignment for an incoming patient. The implementation of randomization procedures without Markov property is deferred to our future work.

\subsubsection{Statistical characteristics for 1:1 randomization procedures}\label{Sec2.1.2}
For 1:1 restricted randomization procedures, a key consideration is a tradeoff between treatment balance and allocation randomness \citep{SverdlovRyeznik2024}. The probability distribution of imbalance $D(j)$ for $j = 1, \ldots, n$ may be an important starting point for obtaining various operating characteristics of a randomization design. The following four measures of imbalance for $j = 1, \ldots, n$ are implemented in Incertus.jl:
\begin{itemize}
    \item $E|D(j)|$ -- expected absolute imbalance at the $j$th step.
    \item $E|D(j)|^2=var\{D(j)\}$ -- variance of imbalance at the $j$th step.
    \item $E\left\{\max_{1\le i\le j}|D(j)|\right\}$ -- expected maximum imbalance over first $j$ allocation steps.
    \item A cumulative average loss \citep{Atkinson2002} over $j$ steps:
    \begin{equation}\label{Eq5}
        L(j) = \frac{1}{j}\sum_{i=1}^j\frac{E|D(i)|^2}{i}.
    \end{equation}
\end{itemize}

As regards lack of randomness, the following four measures for $j = 1, \ldots, n$ are implemented in Incertus.jl:
\begin{itemize}
    \item $EPCG_{conv}(j)$ -- expected proportion of correct guesses over first $j$ allocation steps under the \emph{convergence} guessing strategy \citep{BlackwellHodges1957}. Under this model, the investigator’s guess on an upcoming treatment assignment, $\delta_{i+1}$, depends on the sign of the current imbalance $D(i)$: guess the $(i+1)$st assignment as E, if $D(i)<0$; as C, if $D(i)>0$; or make a random guess with probability 0.5, if $D(i)=0$.
    \item $EPCG_{max}(j)$ -- expected proportion of correct guesses over first $j$ allocation steps under the \emph{maximum probability} guessing strategy: guess the $i$th treatment assignment as E, if $\phi_i > 0.5$; as C, if $\phi_i < 0.5$; or guess at random, if $\phi_i = 0.5$.
    \item $EPDA(j)=\frac{1}{j}\sum_{i=1}^j\Pr(\phi_i=0\;\textrm{or}\;1)$ -- expected proportion of deterministic assignments over first $j$ allocation steps. 
    \item $FI(j)=\frac{1}{j}\sum_{i=1}^jE|\phi_i-0.5|$ -- forcing index \citep{Heritier2005} that takes values on a scale 0--1. We have $FI(j)\equiv 0$ $\forall j$ for CRD (most random design) and $FI(j)=1$ for the permuted blocks of 2, assuming $j$ is even (most balanced design).
\end{itemize}

To quantify the balance–randomness tradeoff at the $j$th allocation step, we consider the Euclidean distance to the origin $(0, 0)$ in the Loss--Forcing Index space \citep{SverdlovRyeznik2024}:
\begin{equation}\label{Eq6}
    G(j) = \sqrt{\{L(j)\}^2+\{FI(j)\}^2},
\end{equation} with lower values of $G(j)$ indicating better balance–randomness tradeoff. A plot of $G(j)$ for $j = 1, \ldots, n$ can provide valuable insights into the dynamic behavior of balance–randomness tradeoff of a chosen randomization procedure, and it can be useful for comparing different randomization procedures.

\subsection{Multi-arm randomization}\label{Sec2.2}
A general restricted randomization procedure for a trial with $K\ge 2$ treatment arms and a fixed target allocation ratio $\rho_1{:}\rho_2{:}\ldots{:}\rho_K$, where $\rho_i\in(0,1)$ and $\sum_{i=1}^K\rho_i=1$ is defined by Equation (\ref{Eq1}). Two important data structures in this context are the $n\times K$ matrix of treatment assignments $\DeltaB_n=(\deltaB_1,\deltaB_2,\ldots,\deltaB_n)^T$ and the $n\times K$ matrix of conditional randomization probabilities $\PhiB_n=(\phiB_1,\phiB_2,\ldots,\phiB_n)^T$ . In the current version of Incertus.jl, we mainly consider randomization procedures with Markov property, i.e., for which the conditional randomization probabilities for an incoming patient depend on the past allocations only through the current treatment numbers, in other words:
\begin{equation}\label{Eq7}
    \phi_{j+1,k} = \Pr(\deltaB_{j+1}=\eB_k|\NB(j)),\;\;j\ge 1,\;k=1,\ldots,K,
\end{equation}where $\NB(j)=(N_1(j),\ldots,N_K(j))^T$ and $N_k(j)=\sum_{i=1}^j\mathbb{1}\{\deltaB_i=\eB_k\}$ is the number of patients assigned to the $k$th treatment after $j$ allocation steps, with $\sum_{k=1}^KN_k(j)=j$.

While most of the multi-arm restricted randomization procedures currently implemented in Incertus.jl have the allocation rules of the form (\ref{Eq7}), some do not (e.g., the drop-the-loser urn design \citep{Ivanova2003} for which the conditional randomization probabilities are determined by the current urn composition rather than by the treatment sample sizes $\NB(j)$). In general, procedures without Markov property require some customized programming, and we defer the broad implementation of such procedures to future work.

\subsubsection{Randomization procedures for multi-arm RCTs}\label{Sec2.2.1}
Below we list the randomization procedures for a $K\ge 2$-arm trial implemented in the current version of Incertus.jl. Note that for a special case of $K = 2$ and $\rho_1 = \rho_2 = 0.5$, a multi-arm randomization procedure may take the form of some 1:1 randomization procedure described in Section~\ref{Sec2.1}.
\begin{itemize}
    \item Complete Randomization Design (CRD) applies multinomial probabilities equal to the target allocation proportions to determine the treatment assignment at each allocation steps: $\phi_{jk}=\rho_k$ for $j=1,\ldots,n$ and $k=1,\ldots,K$.
    \item Procedures that force strict balance for the given sample size: Suppose the target sample sizes (positive integers) for the $K$ treatment arms are $n_1:n_2\ldots:n_K$ with $\sum_{k=1}^Kn_k=n$ (here $n_k=n\rho_k$, rounded to integer if necessary). Two designs that can be used to achieve this allocation exactly are the Random Allocation Rule (Rand) and the Truncated Multinomial Design (TMD) \citep{Rukhin2007}.
    \item Permuted Block Design (PBD): Consider the target allocation ratio $w_1:w_2:\ldots:w_K$, where $w_k$'s are positive integers with the common divisor of 1. Let $W=\sum_{k=1}^Kw_k$ the size of the ``minimal balanced'' set. The PBD performs treatment assignments in blocks of size $bW$, where $b$ is some small positive integer. Within each block, exactly $bw_k$ patients are randomized to the $k$th treatment, $k=1,\ldots,K$.
    \item Randomization procedures that can achieve (approximately) any fixed integer-valued allocation ratio $w_1:w_2:\ldots:w_K$, where $w_k$'s are positive integers with the common divisor of 1, with fewer restrictions than PBD: Block Urn Design (BUD) \citep{ZhaoWeng2011} and Drop-the-Loser Urn Design (DLUD) \citep{Ivanova2003, RyeznikSverdlov2018}.
    \item Randomization procedures that can achieve (approximately) any fixed, possibly irrational-valued allocation proportions $\rho_1{:}\rho_2{:}\ldots{:}\rho_K$, $\rho_k\in(0,1)$ and $\sum_{k=1}^K\rho_k=1$ (in particular, they can be applied with the integer-valued ratios $w_1:w_2:\ldots:w_K$). Three designs that are fit for this purpose are Mass Weighted Urn Design (MWUD) \cite{Zhao2015}, Doubly Adaptive Biased Coin Design (DBCD) \cite{HuZhang2004}, and Maximum Entropy Constrained Balance Randomization (MaxEnt) \citep{Klotz1978, RyeznikSverdlov2018}.
\end{itemize}

\subsubsection{Statistical characteristics for multi-arm randomization procedures}\label{Sec2.2.2}
For a $K\ge 2$-arm trial, by balanced allocation, we mean the achievement of the target allocation ratio for the given sample size. A natural measure of lack of balance after $j$ allocation steps is the Euclidean distance between the observed allocation $\NB(j)=(N_1(j),\ldots,N_K(j))^T$ and the targeted allocation $j\rhoB=(j\rho_1,\ldots,j\rho_K)^T$:
\begin{equation}\label{Eq8}
    d(j)=\sqrt{\sum_{i=1}^K(N_i(j)-j\rho_i)^2}.
\end{equation}Since $N_i(j)$'s are random, $d(j)$ in (\ref{Eq8}) is a random variable, with smaller values indicating more balanced allocation. For $K=2$ and $\rho_1=\rho_2=0.5$, (\ref{Eq8}) is equal to $\frac{1}{\sqrt{2}}|N_1(j)-N_2(j)|=\frac{1}{\sqrt{2}}|D(j)|$. The following measures of lack of balance for $j=1,\ldots,n$ are implemented in Incertus.jl:
\begin{itemize}
    \item $E[d(j)]$ -- expected distance between the observed and the targeted treatment sample sizes at the $j$th step.
    \item $E[d^2(j)]=\sum_{i=1}^Kvar[N_i(j)]$ -- total variance of the treatment sample sizes at the $j$th step.
    \item $E\left\{\max_{1\le i\le j}d(i)\right\}$ -- expected maximum distance between the observed and the targeted treatment sample sizes over first $j$ allocation steps.
    \item A cumulative loss over $j$ allocation steps:
    \begin{equation}\label{Eq9}
        L(j) = \frac{1}{j}\sum_{i=1}^j \frac{E[d^2(i)]}{i}.
    \end{equation}
\end{itemize}To quantify the lack of randomness, we consider the following measures:
\begin{itemize}
    \item $EPCG_{min}(j)$ -- expected proportion of correct guesses over first $j$ allocation steps under the \emph{minimum imbalance} guessing strategy \citep{Zhao2017}.  Under this model, let $\Delta_k(i)=N_k(i)-i\rho_k$ be the difference between the achieved sample size and the ``ideal'' sample size for the $k$th treatment ($k = 1, \ldots, K$) after $i$ allocation steps. Then the investigator guesses the $(i + 1)$st assignment to be treatment $s$ that is currently most underrepresented, i.e., for which $\Delta_s(i)=\min_{1\le k\le K}\Delta_k(i)$ (if several treatments meet this condition, a random guess is made on this set of treatments). 
    \item $EPCG_{max}(j)$ --expected proportion of correct guesses over first $j$ allocation steps under the \emph{maximum probability} guessing strategy \citep{Zhao2017}: guess the $i$th treatment assignment to be treatment $s$ that has the maximum conditional randomization probability, i.e., $\phi_{is}=\max_{1\le k\le K}\phi_{ik}$ (if several treatments meet this condition, a random guess is made on this set of treatments).
    \item $EPDA(j)$ -- expected proportion of deterministic assignments over first $j$ allocation steps. Here a ``deterministic assignment'' at the $i$th allocation step means that some treatment is assigned with probability 1 at the $i$th step. Therefore,
    \begin{equation}\label{Eq10}
        EPDA(j) = \frac{1}{j}E\left[\sum_{i=1}^j\mathbb{1}
        \left\{\exists k:\phi_{ik}=1\;\text{and}\;\forall k'\ne k:\phi_{ik'}=0\right\}\right].
    \end{equation}
    \item Forcing Index (FI) defined as the cumulative average (over $j$ allocation steps) of the expected Euclidean distance from the vector of treatment randomization probabilities and the target allocation proportions:
    \begin{equation}\label{Eq11}
        FI(j) = \frac{1}{j}\sum_{i=1}^jE\sqrt{\sum_{k=1}^K(\phi_{ik}-\rho_k)^2}.
    \end{equation}Small values of $FI(j)$ correspond to more random procedures. For CRD, $FI(j)\equiv 0$ $\forall j$.
\end{itemize}

To quantify balance--randomness tradeoff for a given randomization design, suppose we have $U_B(j)$ and  $U_R(j)$ – the measures of lack of balance and lack of randomness after $j$ allocation steps, normalized to the same scale of 0–1, such that smaller values of $U_B(j)$ indicate better balance, and smaller values of $U_R(j)$ indicate better randomness \citep{SverdlovRyeznik2019}. Then, assuming balance and randomness are equally important, we can define an ``overall'' performance measure
as the distance to the origin (0,0):
\begin{equation}\label{Eq12}
    G(j) = \sqrt{\{U_B(j)\}^2 + \{U_R(j)\}^2},
\end{equation}with lower values of $G(j)$ indicating a better balance--randomness tradeoff.

Finally, to assess the ARP property of a randomization procedure, we obtain the unconditional treatment randomization probabilities at every step. These probabilities can be estimated through Monte Carlo simulations:
\begin{equation}\label{Eq13}
    \pi_{jk} = E(\phi_{jk}) \approx \frac{1}{N_{sim}} \sum_{\ell=1}^{N_{sim}}\phi_{jk}^{\ell},
\end{equation}where $\phi_{jk}^{\ell}$ is the conditional randomization probability for the $k$th treatment at the $j$th allocation step from the $\ell$th simulation run ($j=1,\ldots,n$, $k=1,\ldots,K$, $\ell=1,\ldots,N_{sim}$). A plot of $\pi_{jk}$ vs. allocation step can be very useful -- for a procedure that lacks ARP property one may expect deviations from the target allocation proportions $\rho_k$'s and fluctuations in $\pi_{jk}$ as $j$ varies from 1 to $n$.

\section{Incertus.jl Package}\label{Sec3}
\subsection{Overview}\label{Sec3.1}
Incertus.jl is available at the GitHub repository: \url{https://github.com/yevgenryeznik/Incertus.jl}. To install the package, one should run the command in Julia REPL:

\noindent \texttt{] add "https://github.com/yevgenryeznik/Incertus.jl"}

The package documentation is available at \url{https://yevgenryeznik.github.io/Incertus.jl/build/}.

Tables~\ref{tab:T1}--\ref{tab:T2} summarize the functions and methods available in Incertus.jl.
\begin{table}
	\caption{Functions for initializing different randomization procedures in Incertus.jl package}
	\centering
	\begin{tabular}{ll}
    \hline
		\multicolumn{2}{c}{Functions for initializing 1:1 randomization procedures}\\
    \hline
		\texttt{CRD()} & initializes a complete randomization design \\
		\texttt{RAND(n)} & initializes a random allocation rule; sample size equals to $n$ \\
		\texttt{TBD(n)} & initializes a truncated binomial design; sample size equals to $n$ \\
		\texttt{PBD(b)} & initializes a permuted block design with a block size equal to $2b$ \\  
		\texttt{BSD(b)} & initializes a big stick design with MTI parameter equal to $b$ \\  
		\texttt{BCDWIT(p,b)} & initializes biased coin design with imbalance tolerance; biased coin probability = $p$; MTI parameter = $b$ \\  
		\texttt{EUD(b)} & initializes Ehrenfest urn design with MTI parameter equal to $b$ \\  
		\texttt{BUD($\lambda$)} & initializes block urn design with a parameter equal to $\lambda$ \\ 
		\texttt{EBCD(p)} & initializes Efron's biased coin design; biased coin probability = $p$ \\  
		\texttt{ABCD(p)} & initializes adjustable biased coin design with a parameter equal to $a$ \\
		\texttt{GBCD($\gamma$)} & initializes generalized biased coin design with a parameter equal to $\gamma$ \\
		\texttt{BBCD($\gamma$)} & initializes the Bayesian biased coin design with a parameter equal to $\gamma$ \\
    \hline
		\multicolumn{2}{c}{Functions for initializing multi-arm randomization procedures}\\    
		\multicolumn{2}{c}{(an input parameter vector $\wB$ specifies target allocation)}\\ 
    \hline
        \texttt{CRD(w)} & initializes a complete randomization design \\
		\texttt{RAND(w,n)} & initializes a random allocation rule; sample size equals to $n$ \\
		\texttt{TBD(w,n)} & initializes a truncated multinomial design; sample size equals to $n$ \\
		\texttt{PBD(w,b)} & initializes a permuted block design; block size equals to $b(w_1+\ldots+w_K)$ \\
		\texttt{BUD(w,$\lambda$)} & initializes block urn design with a parameter equal to $\lambda$ \\ 
		\texttt{MWUD(w,$\alpha$)} & initializes mass weighted urn design with a parameter equal to $\alpha$ \\ 
		\texttt{DLUD(w,a)} & initializes drop-the-loser urn design with a parameter equal to $a$ \\ 
		\texttt{DBCD(w,$\gamma$)} & initializes doubly adaptive biased coin design with a parameter equal to $\gamma$ \\ 
		\texttt{MaxEnt(w,$\eta$)} & initializes MaxEnt design with a parameter equal to $\eta$ \\ 
    \hline
	\end{tabular}\label{tab:T1}
\end{table}

\begin{table}
	\caption{Methods implemented in Incertus.jl package}
	\centering
	\begin{tabular}{ll}
    \hline
		\multicolumn{2}{c}{Simulation of randomization procedures}\\
    \hline
		\texttt{simulate(rnd,nsbj,nsim,seed)} & simulates a randomization. Here \texttt{rnd} represents either a single randomization \\
		& procedure or a set of procedures to be simulated; \texttt{nsbj} is a sample size; \\
        & \texttt{nsim} is the number of simulations; and \texttt{seed} is used for initializing a random \\
        & number generator (a default is set to 314159) \\  
    \hline
		\multicolumn{2}{c}{Calculation of operating characteristics}\\
        \multicolumn{2}{c}{(an input parameter \texttt{sr} is the output of the \texttt{simulate} function, representing the simulation results)}\\
    \hline
		\multicolumn{2}{c}{Balance}\\
    \hline
    \texttt{calc\_final\_imb(sr)} & calculates the distribution of the final imbalance \\
    \texttt{calc\_expected\_abs\_imb(sr)} & calculates expected absolute imbalance vs. allocation step \\    
    \texttt{calc\_variance\_of\_imb(sr)} & calculates variance of imbalance vs. allocation step \\  
    \texttt{calc\_expected\_max\_abs\_imb(sr)} & calculates expected maximum absolute imbalance vs. allocation step\\  
    \texttt{calc\_cummean\_loss(sr)} & calculates cumulative average loss vs. allocation step\\  
    \hline
		\multicolumn{2}{c}{Randomness}\\
    \hline
    \texttt{calc\_cummean\_epcg(sr,gs)} & calculates cumulative averages of expected proportions of correct guesses vs. \\
    & allocation step; \texttt{gs} parameter accepts two values: “C” (converging guessing strategy) \\
    & or “MP” (maximum probability guessing strategy)\\ 
    \texttt{calc\_cummean\_pda(sr)} & calculates cumulative averages of the proportions of deterministic assignments vs. \\
    &  allocation step \\ 
    \texttt{calc\_fi(sr)} & calculates forcing index vs. allocation step \\
    \hline
		\multicolumn{2}{c}{Balance-randomness tradeoff and the allocation ratio preserving (ARP) property}\\
    \hline    
    \texttt{calc\_brt(sr)} & calculates balance-randomness tradeoff vs. allocation step\\    
    \texttt{eval\_arp(sr)} & evaluates ARP property, i.e., calculates unconditional allocation probabilities\\     
    \hline
		\multicolumn{2}{c}{Visualization of operating characteristics}\\
        \multicolumn{2}{c}{(\texttt{op} is a data frame with corresponding operating characteristics;  \texttt{brt} is an output of \texttt{calc\_brt} function; }\\ 
        \multicolumn{2}{c}{\texttt{arp} is an output of \texttt{eval\_arp} function; \texttt{final\_imb} is an output of \texttt{calc\_final\_imb} function; }\\
        \multicolumn{2}{c}{\texttt{kwargs} is a set of key words that can be accepted by the visualization functions} \\
    \hline
    \texttt{violin(final\_imb; kwargs...)} & visualize a distribution of the final imbalance as a violin plot\\
    \texttt{heatmap(brt; kwargs...)} & visualize balance-randomness trade-off as a heatmap plot\\
    \texttt{plot(op; kwargs...)} & visualize simulated operating characteristics\\
    \texttt{plot(arp; kwargs...)} & visualize simulated unconditional allocation probabilities\\
    \hline 
	\end{tabular}\label{tab:T2}
\end{table}

\subsection{Examples: Two-arm equal randomization}\label{Sec3.2}
\subsubsection{Generating a randomization sequence for a 1:1 RCT}\label{Sec3.2.1}

The function \texttt{simulate} is a core function for sequentially generating treatment randomization probabilities and treatment assignments for the chosen randomization method(s). Suppose we want to generate a randomization sequence for a 1:1 RCT with $n=8$. We consider four randomization procedures:
\begin{itemize}
    \item PBD with blocks of size $2b=4$ -- \texttt{PBD(2)}
    \item Random allocation rule -- \texttt{RAND(8)}
    \item BSD with $b=3$ -- \texttt{BSD(3)}
    \item Efron's BCD with $p=2/3$ -- \texttt{EBCD(2/3)}
\end{itemize}The following code is used:

\noindent \texttt{using Incertus}\\ \\
\noindent \texttt{\# sample size}\\
\noindent \texttt{nsbj = 8;}\\ \\
\noindent \texttt{\# number of simulations is set to 1}\\
\noindent \texttt{nsim = 1;}\\ \\
\noindent \texttt{\# randomization procedures to be simulated}\\
\noindent \texttt{rnd = [PBD(2), RAND(nsbj), BSD(3), EBCD(2//3)];}\\ \\
\noindent \texttt{\# labels}\\
\noindent \texttt{label\_rnd = [label(item) for item in rnd]}\\ \\
\noindent \texttt{\# simulation run}\\
\noindent \texttt{sr = simulate(rnd, nsbj, nsim);}\\

The function \texttt{simulate} takes as the input the set of four specified randomization procedures \texttt{(rnd = [PBD(2), RAND(nsbj), BSD(3), EBCD(2//3)])}, the sample size \texttt{(nsbj = 8)}, and the number of simulations (\texttt{nsim = 1}, which indicates that we are interested in simulating a single sequence for each randomization method). Also, it is possible to provide a seed for a random number generator as a fourth input argument, which is set to 314159 by default. The output object \texttt{sr} contains four data structures corresponding to four implemented randomization procedures. For example, \texttt{sr[1]} has the output for \texttt{PBD(2)}, \texttt{sr[2]} has the output for \texttt{RAND}, etc. Then the $8\times 2$ matrix of treatment assignments and the $8\times 2$ matrix of conditional randomization probabilities for \texttt{PBD(2)} can be obtained as \texttt{sr[1].trt} and \texttt{sr[1].prb}, respectively. Similar commands can be used to extract the corresponding data structures for the other three randomization procedures.

We can also create data frames combining treatment E assignments (\texttt{trt\_df}) and conditional randomization probabilities for treatment E (\texttt{prb\_df}) of the four considered randomization procedures using the following code:

\noindent \texttt{using DataFrames}\\ \\
\noindent \texttt{trt = hcat([sr[i].trt[:, 1, 1] for i in eachindex(sr)]...);}\\
\noindent \texttt{trt\_df = DataFrame(trt, label\_rnd)}\\ \\
\noindent \texttt{prb = hcat([sr[i].prb[:, 1, 1] for i in eachindex(sr)]...);}\\
\noindent \texttt{prb\_df = DataFrame(prb, label\_rnd)}\\

\begin{figure}
	\centering
    \includegraphics[width=1\textwidth]{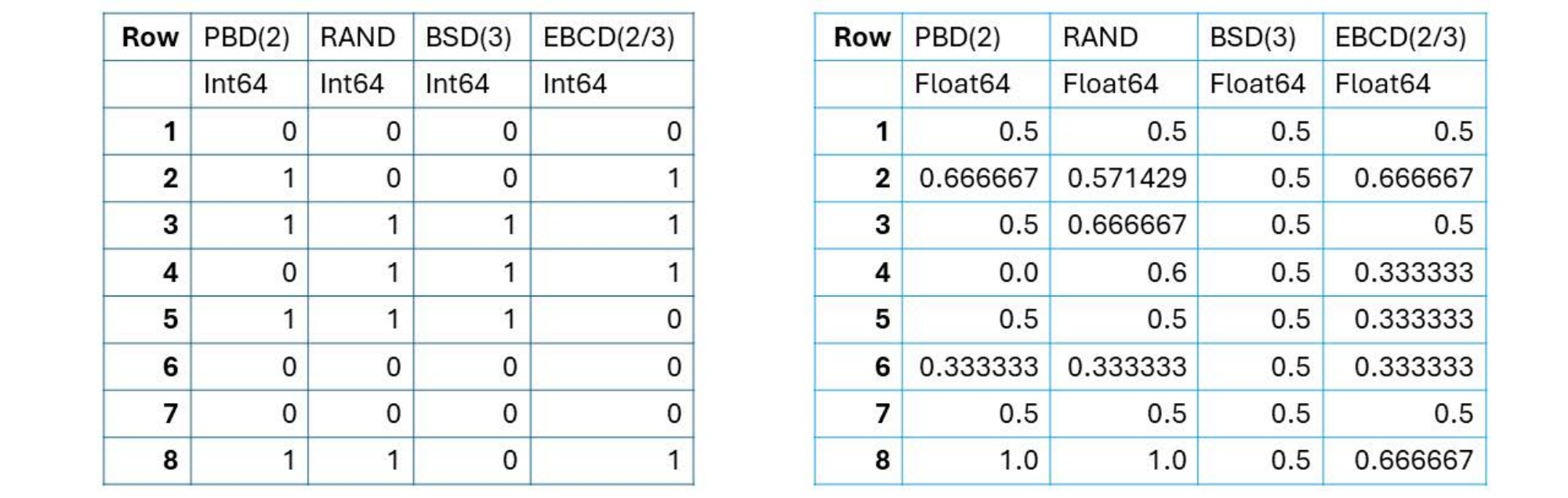}
	\caption{Data frames of treatment assignments (left) and conditional randomization probabilities (right) of four randomization designs -- \texttt{PBD(2)}, \texttt{RAND}, \texttt{BSD(3)}, and \texttt{EBCD(2/3)} -- for a 1:1 RCT with $n=8$.}
	\label{Figure1}
\end{figure}

The resulting data frames are displayed in Figure~\ref{Figure1}. One can see that the four randomization procedures have different vectors of conditional randomization probabilities (displayed as columns
of the right table) and result in different treatment allocation sequences (displayed as columns of the left table). See Supplemental Appendix, “Simulation example 1” for details.

\subsubsection{Assessing statistical properties of 1:1 randomization designs through Monte Carlo simulation}\label{Sec3.2.2}
Monte Carlo simulations provide a useful tool for evaluating various randomization designs and selecting one that is most fit for purpose for the given trial parameters. Suppose we want to compare seven randomization designs for a 1:1 RCT with $n = 40$ patients with respect to the criteria of balance and randomness, as described in Section~\ref{Sec2.1.2}. The designs are:
\begin{itemize}
    \item Complete randomization design -- \texttt{CRD}
    \item PBD with blocks of size $2b = 2$ -- \texttt{PBD(1)}
    \item Random allocation rule -- \texttt{RAND}
    \item Truncated binomial design -- \texttt{TBD}
    \item Efron's BCD with $p=2/3$ -- \texttt{EBCD(2/3)}
    \item Adjustable BCD with $a=2$ -- \texttt{ABCD(2)}
\end{itemize}A Monte Carlo simulation (10,000 runs) is performed using the code:

\noindent \texttt{using Incertus}\\ \\
\noindent \texttt{\# sample size}\\
\noindent \texttt{nsbj = 40;}\\ \\
\noindent \texttt{\# number of simulations is set to 1}\\
\noindent \texttt{nsim = 10000;}\\ \\
\noindent \texttt{\# randomization procedures to be simulated}\\
\noindent \texttt{rnd = [CRD(), PBD(1), RAND(nsbj), TBD(nsbj), BSD(3), EBCD(2//3), ABCD(2);}\\ \\
\noindent \texttt{\# simulation run}\\
\noindent \texttt{sr = simulate(rnd, nsbj, nsim);}\\

Based on the output object, \texttt{sr}, the selected design operating characteristics can be calculated and visualized using Julia graphical tools using the following code:

\noindent \texttt{\# calculating cumulative average loss vs. allocation step}\\
\noindent \texttt{cummean\_loss = calc\_cummean\_loss(sr);}\\ 
\noindent \texttt{\# creating a plot of cumulative average losses over the first allocation steps}\\
\noindent \texttt{plot(cummean\_loss, ylabel = "cumulative average loss", size = (800, 600))}\\ \\
\noindent \texttt{\# calculating forcing index vs. allocation step}\\
\noindent \texttt{fi = calc\_fi(sr);}\\ 
\noindent \texttt{\# creating a plot of forcing indices vs. allocation step}\\
\noindent \texttt{plot(fi, ylabel = "forcing index", size = (800, 600))}\\ \\
\noindent \texttt{\# Evaluating balance-randomness tradeoff vs. allocation step}\\
\noindent \texttt{brt = calc\_brt(sr);}\\ 
\noindent \texttt{\# creating a heatmap plot of balance--randomness tradeoff vs. allocation step}\\
\noindent \texttt{heatmap(brt, size = (800, 600))}\\ \\
\noindent \texttt{\# calculating final imbalance}\\
\noindent \texttt{final\_imb = calc\_final\_imb(sr);}\\ 
\noindent \texttt{\# creating a violin plot of final imbalances}\\
\noindent \texttt{violin(final\_imb, size = (800, 600))}\\

\begin{figure}[h]
	\centering
    \includegraphics[width=1\textwidth]{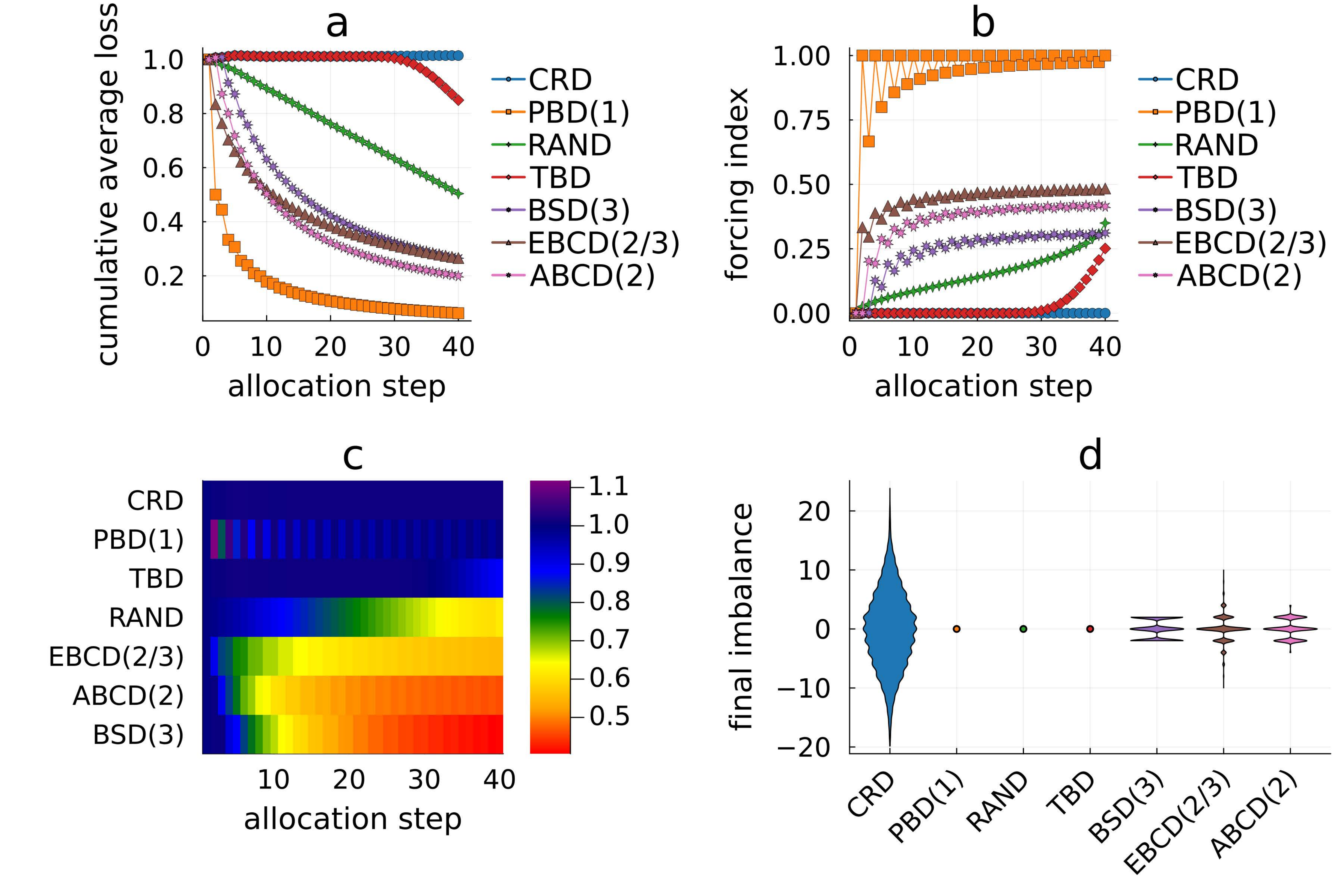}
	\caption{Selected operating characteristics of seven randomization designs -- \texttt{CRD}, \texttt{PBD(1)}, \texttt{RAND}, \texttt{TBD}, \texttt{BSD(3)}, \texttt{EBCD(2/3)}, and \texttt{ABCD(2)} -- for a 1:1 RCT with $n=40$.}
	\label{Figure2}
\end{figure}

Figure~\ref{Figure2}(a--d) displays, respectively, the plots of cumulative average loss, forcing index, balance-randomness tradeoff heatmap plot, and the violin plot of final imbalance. We can see that among the seven considered designs in this scenario, \texttt{BSD(3)} provides, overall, the best tradeoff between balance and randomness. The full report of this simulation is available in the Supplemental Appendix, “Simulation example 2”.

\subsubsection{Calibrating 1:1 randomization design through Monte Carlo simulation}\label{Sec3.2.3}
Most restricted randomization designs involve a parameter that can be fine-tuned to determine a design with a desirable balance-randomness tradeoff. For example, consider the Bayesian Biased Coin Design (BBCD) with five different values of the parameter $\gamma = 0.01; 0.05; 0.1; 0.2;$ and 1, for a 1:1 RCT with $n = 40$. We ran 10,000 Monte Carlo simulations to obtain the operating characteristics of these designs. The Julia code and the full report of simulation results are available in the Supplemental Appendix, “Simulation example 3". Figure~\ref{Figure3} below shows a heatmap plot of the balance--randomness tradeoff metric. One can see that $\gamma = 0.05$ is an ``optimal'' choice -- it results in increasingly smaller values of the metric over allocation steps, the smallest among the five considered designs.

\begin{figure}[h]
	\centering
    \includegraphics[width=1\textwidth]{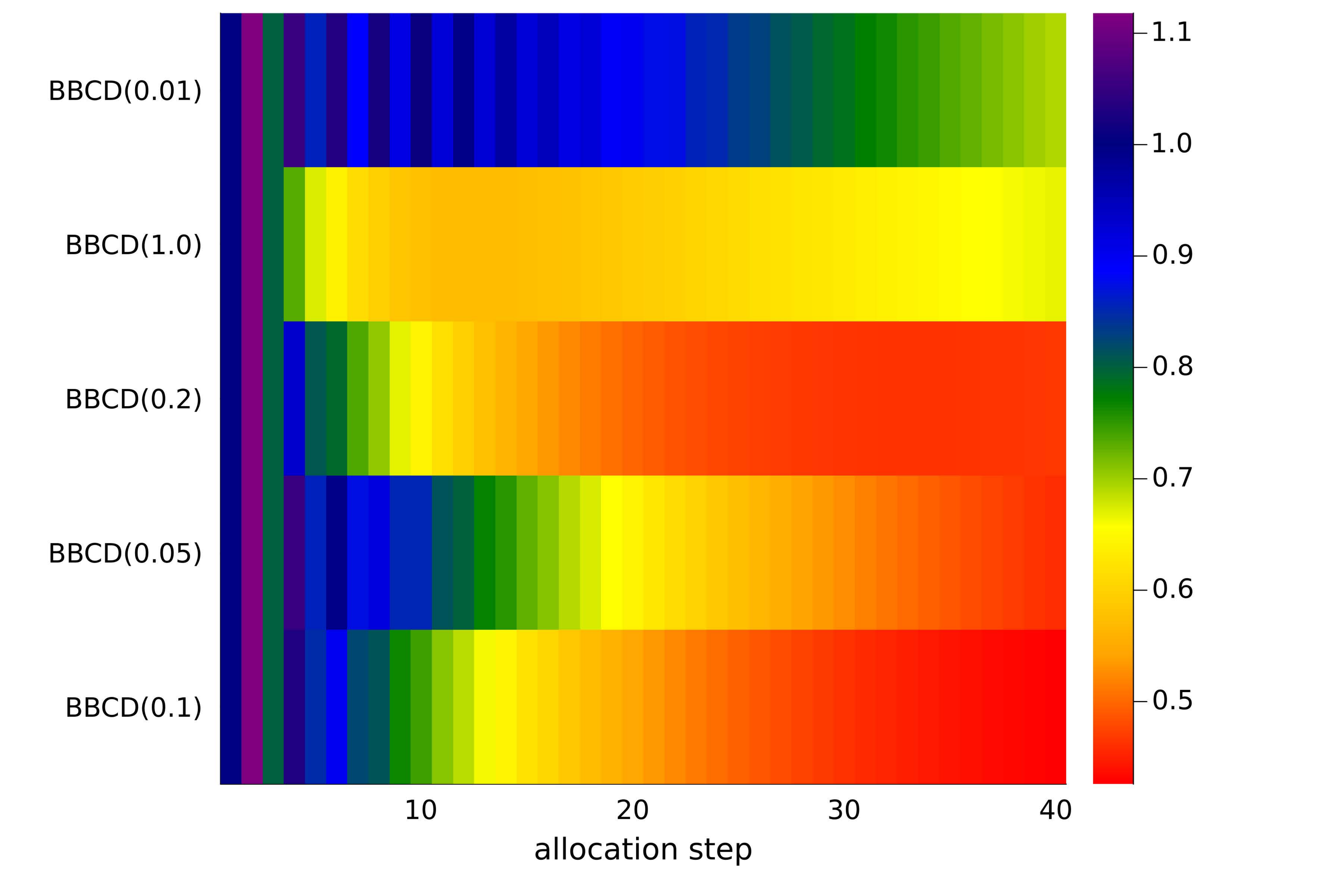}
	\caption{Heatmap plot of balance-randomness tradeoff for the Bayesian biased coin design (\texttt{BBCD}) with five different values of the parameter $\gamma = 0.01; 0.05; 0.1; 0.2;$ and 1, for a 1:1 RCT with $n = 40$.}
	\label{Figure3}
\end{figure}

\subsection{Examples: Multi-arm equal randomization}\label{Sec3.3}
\subsubsection{Generating a randomization sequence for a 4:3:2:1 RCT}\label{Sec3.3.1}

Here, we demonstrate how to generate a randomization sequence and randomization probabilities for a multi-arm RCT, targeting 4:3:2:1 allocation, with $n = 10$. We consider two randomization designs: random allocation rule (\texttt{RAND}) and truncated multinomial design (\texttt{TMD}). The following code is used:

\noindent \texttt{using Incertus}\\ \\
\noindent \texttt{\# sample size}\\
\noindent \texttt{nsbj = 10;}\\ \\
\noindent \texttt{\# number of simulations is set to 1}\\
\noindent \texttt{nsim = 1;}\\ \\
\noindent \texttt{\# target allocation}\\
\noindent \texttt{w = [4, 3, 2, 1];}\\ \\
\noindent \texttt{\# randomization procedures to be simulated}\\
\noindent \texttt{rnd = [RAND(w, nsbj), TMD(w, nsbj)];}\\ \\
\noindent \texttt{\# labels}\\
\noindent \texttt{label\_rnd = [label(item) for item in rnd]}\\ \\
\noindent \texttt{\# simulation run}\\
\noindent \texttt{sr = simulate(rnd, nsbj, nsim);}\\

The output object \texttt{sr} contains two data structures: \texttt{sr[1]} for \texttt{RAND} and \texttt{sr[2]} for \texttt{TMD}. For
\texttt{RAND}, the $10\times 4$ matrix of treatment assignments is retrieved as \texttt{sr[1].trt}, and the $10\times 4$ matrix of conditional randomization probabilities is retrieved as \texttt{sr[1].prb}. Likewise, for \texttt{TMD}, the corresponding matrices are retrieved as \texttt{sr[2].trt} and \texttt{sr[2].prb}. See Supplemental Appendix, ``Simulation example 5'' for details.

\subsubsection{Assessing statistical properties of 4:3:2:1 randomization designs through Monte Carlo simulation}\label{Sec3.3.2}
Consider a four-arm RCT with $n = 40$ patients, targeting 4:3:2:1 allocation. We are interested in comparing seven randomization designs with respect to the criteria of balance and randomness, as described in Section~\ref{Sec2.2.2}. The designs are:
\begin{itemize}
    \item Complete randomization design -- \texttt{CRD}
    \item PBD with a block size $bW = 1\cdot(4+3+2+1)=10$ -- \texttt{PBD(1)}
    \item Block urn design with $\lambda=2$ -- \texttt{BUD(2)}
    \item Random allocation rule -- \texttt{RAND}
    \item Truncated multinomial design -- \texttt{TMD}
    \item Drop-the-loser urn design with $a=2$ -- \texttt{DLUD(2)}
    \item Mass weighted urn design with $\alpha=2$ -- \texttt{MWUD(2)}
\end{itemize}A Monte Carlo simulation (10,000 runs) is performed using the code:

\noindent \texttt{using Incertus}\\ \\
\noindent \texttt{\# sample size}\\
\noindent \texttt{nsbj = 40;}\\ \\
\noindent \texttt{\# number of simulations}\\
\noindent \texttt{nsim = 10000;}\\ \\
\noindent \texttt{\# target allocation}\\
\noindent \texttt{w = [4, 3, 2, 1];}\\ \\
\noindent \texttt{\# randomization procedures to be simulated}\\
\noindent \texttt{rnd = [CRD(w), PBD(w, 1), BUD(w, 2), RAND(w, nsbj), TMD(w, nsbj), DLUD(w, 2), MWUD(w, 2)];}\\ \\
\noindent \texttt{\# simulation run}\\
\noindent \texttt{sr = simulate(rnd, nsbj, nsim);}\\

The design operating characteristics are calculated based on the output object \texttt{sr}, and they are visualized using Julia graphical tools; see Supplemental Appendix, “Simulation example 6”.

\subsubsection{Calibrating a 4:3:2:1 randomization design through Monte Carlo simulation}\label{Sec3.3.3}
\begin{figure}[h]
	\centering
    \includegraphics[width=1\textwidth]{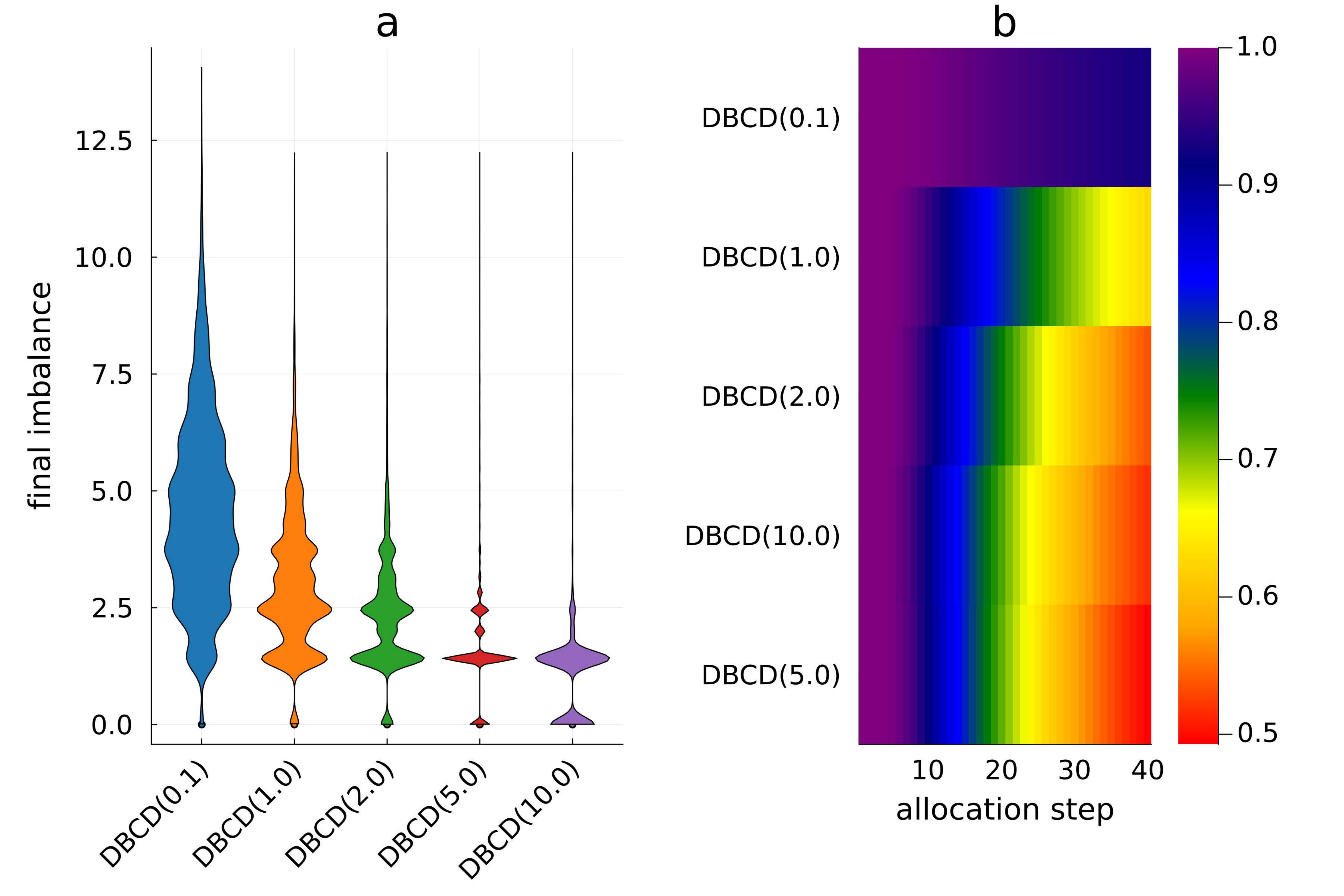}
	\caption{(a) Violin plot of the final imbalance; and (b) Heatmap plot of balance--randomness tradeoff -- for DBCD with five different values of the parameter $\gamma = 0.01; 1; 2; 5;$ and 10 -- for a 4:3:2:1 RCT with $n=40$.}
	\label{Figure4}
\end{figure}

\begin{figure}[h]
	\centering
    \includegraphics[width=1\textwidth]{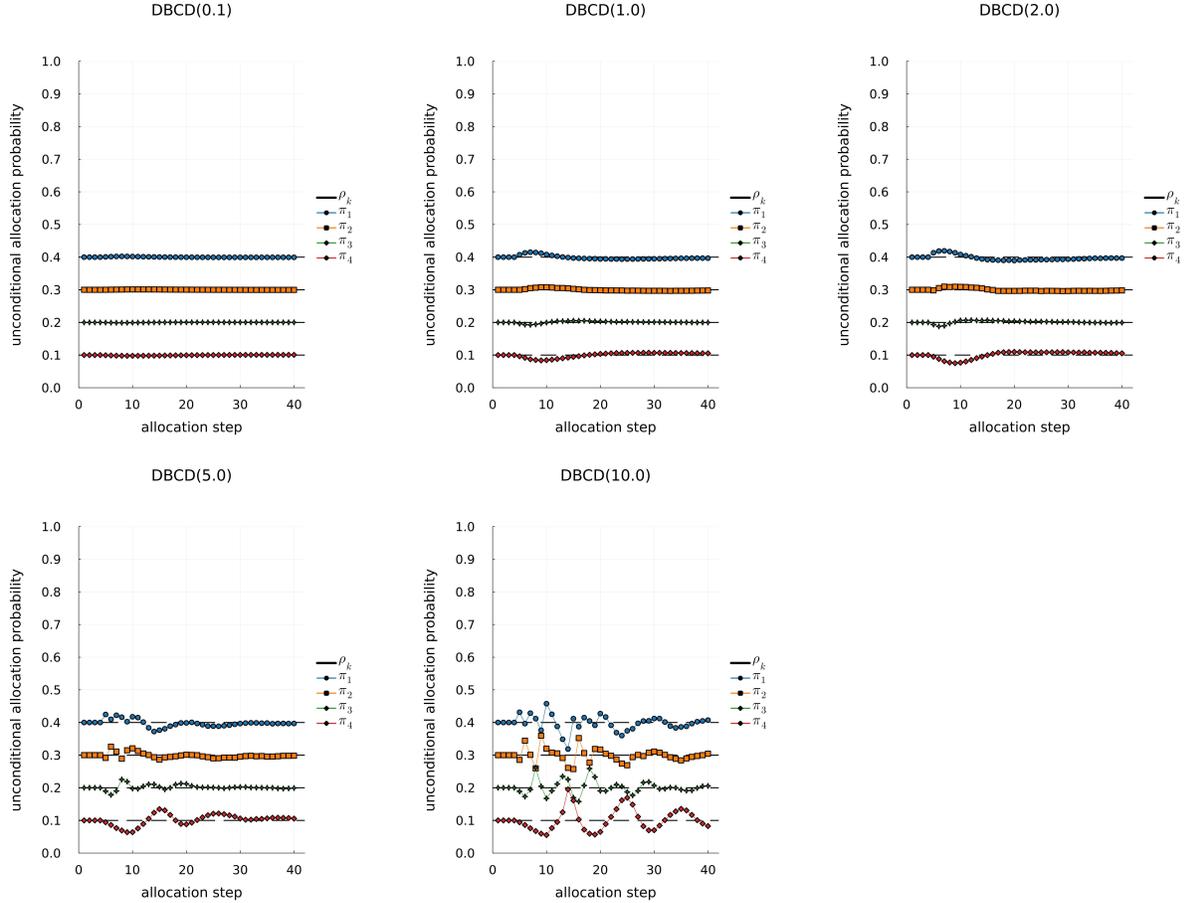}
	\caption{Simulated unconditional randomization probabilities for the four treatment arms -- for DBCD with five different values of the parameter $\gamma = 0.01; 1; 2; 5;$ and 10 -- for a 4:3:2:1 RCT with $n=40$.}
	\label{Figure5}
\end{figure}

Suppose an experimenter decides to use a doubly adaptive biased coin design (DBCD) for a 4:3:2:1 RCT with $n = 40$ patients. Our goal is to find a value of the parameter $\gamma > 0$ that would result in a procedure with a sensible balance-randomness tradeoff. For illustrative purposes, let us consider DBCD with $\gamma = 0.01; 1; 2; 5;$ and 10. We ran 10,000 Monte Carlo simulations to obtain the operating characteristics of these designs. The Julia code and the full report of simulation results are available in the Supplemental Appendix, “Simulation example 7”. Based on these simulations, Figure~\ref{Figure4} shows the violin plot of the final imbalance (left plot) and the heatmap plot of the balance--randomness tradeoff metric (right plot). Figure~\ref{Figure5} shows the plot of simulated unconditional randomization probabilities for the four treatment arms. From Figure~\ref{Figure5}, the DBCD does not have the ARP property because there are some fluctuations (deviations from the target allocation proportions) in the unconditional probabilities; these fluctuations are negligible for $\gamma = 0.1$, and they are more pronounced for larger values of $\gamma$.

\section{Discussion and future work}\label{Sec4}
Incertus.jl is a Julia software package for generating randomization sequences for clinical trials with two or more treatment arms, with equal or unequal allocation ratios. The package is highly efficient computationally and it can be invoked in R. It has a utility for performing simulation studies of selected randomization procedures and assessing their statistical properties, which could help the investigators select a randomization design that fits for purpose of their study. While our examples in the current paper concerned relatively simple cases, the software provides ``building blocks'' for developing randomization designs for more complex studies -- e.g., trials with stratified randomization \citep{Sverdlov2024}, platform trials \citep{APT2019}, etc.

It is worth highlighting some opportunities for extension and future work. First, adding more randomization procedures to the package would be useful. The procedures with Markov property (whose conditional randomization probabilities have analytical expressions as functions of current treatment sample sizes) can be easily programmed following the existing template in Incertus.jl. For the procedures that do not have Markov property, such as the maximal procedure of \cite{Berger2003}, some more customized approach to programming will be warranted.

Second, it will be important to enrich the package with some functions that would enable characterizing reference sets of various randomization procedures, i.e., the sets of all possible randomization sequences and their probabilities. Such an option is currently available in the randomizeR package \citep{Uschner2018} for 1:1 RCTs, and it would be also useful to implement for more general settings, such as multi-arm RCTs with equal or unequal allocation ratios, which are within the scope of Incertus.jl. Having a tool that enables efficient sampling of randomization sequences from the reference set of a given randomization procedure would allow investigators to construct randomization-based tests \citep{Rosenberger2019, Proschan2019} that complement standard likelihood-based tests.

Finally, would be beneficial to develop a user-friendly interface for Incertus.jl. Since the package was developed in Julia, a natural way is to use Julia functionality for building such an interface. At the same time, many recent statistical software applications are developed as R Shiny tools. Finding a way to cast Incertus.jl as an R Shiny tool would be an interesting future research project.

\newpage
\bibliographystyle{unsrtnat}
\bibliography{template.bbl}  






\end{document}